\documentstyle[prl,aps,preprint]{revtex}

\let\ssection=\section
\renewcommand{\section}{\setcounter{equation}{0}\ssection}

\begin{document}
\draft

\title{On cosmic rotation\footnote{Published in: {\sl ``Gravity, 
Particles and Space-Time''}, eds. P. Pronin and G. Sardanashvily (World 
Scientific: Singapore, 1996), pp. 421-439.}}

\author{Vladimir A. Korotky}

\address{Kaluga Division of Moscow N.Bauman Technical University,
Bazhenova 4, 248600 Kaluga, Russia}

\author{Yuri N.~Obukhov\footnote{
On leave from:  Department of Theoretical Physics, Moscow State University,
117234 Moscow, Russia}}
 
\address{Institute for Theoretical Physics, University of Cologne,
D--50923 K\"oln, Germany}

\maketitle

\begin{abstract}
\medskip
We overview our recent studies of cosmological models with expansion and 
global rotation. Problems of the early rotating models are discussed, and 
the class of new viable cosmologies is described in detail. Particular 
attention is paid to the observational effects of the cosmic rotation.
\end{abstract}

\section{Introduction}

This paper is dedicated to the memory of Professor Dmitri D. Ivanenko who was
deeply interested in the problem of universal rotation and made essential
contributions to this subject. For the first time attention to cosmological 
models with rotation was drawn in 1946 by George Gamov \cite{gamov}
(although we should mention also the earlier work of Lanczos \cite{lanc}).
Soon after this, K. G\"odel \cite{godel} had suggested to describe cosmic
rotation with the help of a spacetime metric of the form
\begin{equation}
ds^2 = a^2 (dt^2 - 2e^{x}dtdy + {1\over 2}e^{2x}dy^2 - dx^2 - dz^2).\label{gd}
\end{equation}
Matter in this model is dust with the energy density $\varepsilon$, and the
cosmological constant $\Lambda$ is nontrivial and negative (i.e. its sign
is opposite to that introduced by Einstein). The angular velocity $\omega$
of the cosmic rotation in (\ref{gd}) is given by $\omega^2 = {1\over 2a^2}= 
4\pi G\varepsilon = - \Lambda$. For many years this model became a
theoretical ``laboratory" for the study of rotating cosmologies. As compared
to the G\"odel's world, the model suggested earlier by Lanczos appears to be
less physical in that it describes a universe as a rigidly rotating 
dust cylinder of infinite radius. Dust density in this solution (later 
rederived by van Stockum \cite{van}) diverges at radial infinity. 

Using the G\"odel model one can clearly understand the idea of the cosmic
rotation of matter in the universe \cite{abs}: let us consider a particle
with the initial velocity (in the comoving coordinates) $\{\dot{t}=1, \dot{x}=
\beta, \dot{y}=0, \dot{z}=0\}$ which starts moving from $\{t=x=y=z=0\}$.
Straightforward analysis shows that such a particle deviates from the initial
$x$-axis direction in the rotating metric.

We have no intention of giving a complete review of all the cosmological
models with rotation. Instead we present here our understanding of the 
main problems of the cosmic rotation, and explain how, in our opinion, these
problems can be solved. 

\section{Problems of cosmic rotation}

It is worthwhile to notice that along with the constant deep interest to the
rotating cosmologies, historical development revealed several problems which
were considered by the majority of relativists as the arguments against the
models with nontrivial cosmic rotation. It seems useful to list them here.
In the next sections we will demonstrate that it is possible to solve all
these problems within the framework of wide class of viable rotating
cosmological models.

\subsection{First problem: causality}

G\"odel himself proved the existence of closed time-like curves in the 
metric (\ref{gd}). This was immediately recognized as an unphysical property
because it violates the causal structure of space-time. Considerable efforts
were thus focused on deriving completely causal rotating cosmologies. In his
last work devoted to rotation, G\"odel without proof mentions the possibility
of positive solution of the causality problem \cite{godel2}. First explicit
solutions were reported later \cite{maitra,osch}. Maitra \cite{maitra}
formulated a simple criterium for the existence of closed time-like curves in
rotating metrics. 

\subsection{Second problem: expansion}

Apparent expansion of the universe is usually related to the 
fact of the red shift in the spectra of distant galaxies, and thus all the
standard cosmological models are necessarily non-stationary. However, it was
immediately noticed that it is impossible to combine pure rotation and
expansion in a solution of the general relativity field equations for
a simple physical matter source. Some solutions are known 
\cite{silk,mat,bac,nr,rt,ros,aga,bs,ell,gron,gronsol,ikpan}
which describe certain stages of the universe's evolution, but the complete
cosmological scenario for a rotating world was not available.  

\subsection{Third problem: microwave background radiation}

Discovery of the microwave background radiation has revealed the remarkable
fact that its temperature distribution is isotropic to a very high degree.
This fact was for a long time considered as a serious argument for isotropic
cosmological models and was used for obtaining estimates on the possible
anisotropies which could take place on the early stages of the universe's
evolution. In particular, homogeneous anisotropic rotating cosmologies were
analysed in \cite{haw,col,bjs}, and strong upper limits on the value of the
cosmic rotation were obtained. Plainly speaking, these numerical estimates
did not leave any chance for rotation to be a significant factor in cosmology.

\subsection{Fourth problem: observations}

The last but not least problem is the lack of direct observational evidence 
for the cosmic rotation. Attempting at its experimental discovery one should
study possible systematic irregularities in angular (ideally, over the whole
celestial sphere) distributions of visible physical properties of sources 
located at cosmological distances. Unfortunately, although a lot of data is 
already potentially accumulated in various astrophysical catalogues, no
complete analysis was made in search of the global cosmic rotation. Partially 
this was explained by the insufficient theoretical study of observational 
cosmology with rotation. To our knowledge, till the recent time there were 
only few theoretical predictions concerning the possible manifestations of 
the cosmic rotation, mainly these were the above mentioned estimates of MBR 
anisotropies \cite{haw,col,bjs} (also of the $X$-ray background anisotropies 
\cite{wolfe}) and the number counts tests analyses  
\cite{lanc,godel2,wesson,mavr,fen}. Few purely empirical studies (without 
constructing general relativistic models) of the angular distributions of
astrophysical data are available which interpreted the observed systematic 
irregularities as the possible effects of rotation, 
\cite{birch1,birch2,conw,andr1,andr2}.

\section{Class of shear-free cosmological models with rotation and expansion}

In \cite{banach,jetp} we considered a wide class of viable cosmological 
models with expansion and rotation. Let us describe it here briefly. 
Denoting $x^0 = t$ as the cosmological time and $x^i , i=1,2,3$ as three
spatial coordinates, we write the space-time interval in the form
\begin{equation}
ds^2 = dt^2 - 2Rn_{i}dx^{i}dt - R^{2}\gamma_{ij}dx^{i}dx^{j},\label{met0}
\end{equation}
where $R=R(t)$ is the scale factor, and
\begin{equation}
n_{i}=\nu_{a}e_{i}^{(a)},\quad \gamma_{ij}=\beta_{ab}e_{i}^{(a)}e_{j}^{(b)}.
\label{ng}
\end{equation}
Here ($a,b=1,2,3$) $\nu_{a},\beta_{ab}$ are constant coefficients, while
\begin{equation}
e^{(a)}= e_{i}^{(a)}(x)dx^i \label{ea}
\end{equation}
are the invariant $1$--forms with respect to the action of a three-parameter
group of motion which is admitted by the space-time (\ref{met0}). We assume
that this group acts simply-transitively on the spatial ($t=const$) 
hypersurfaces. It is well known that there exist 9 types of such manifolds,
classified according to the Killing vectors $\xi_{(a)}$ and their commutators
$[\xi_{(a)},\xi_{(b)}]=C^{c}{}_{ab}\xi_{(c)}$. Invariant forms (\ref{ea})
solve the Lie equations ${\cal L}_{\xi_{(b)}}e^{(a)}=0$ for each Bianchi
type, so that models (\ref{met0}) are spatially homogeneous. 

Kinematical characteristics of (\ref{met0}) are as follows: volume expansion
is 
\begin{equation}
\vartheta = 3{\dot{R}\over R},\label{expan}
\end{equation}
nontrivial components of vorticity tensor are
\begin{equation}
\omega_{ij}=-{R\over 2}\hat{C}^{k}{}_{ij}n_{k},\label{rot}
\end{equation}
and shear tensor is trivial,
\begin{equation}
\sigma_{\mu\nu}=0.\label{shear}
\end{equation}
Hereafter the dot ($\dot{}$) denotes derivative with respect to the 
cosmological time coordinate $t$. Tensor $\hat{C}^{k}{}_{ij}=
e^{k}_{(a)}(\partial_{i}e^{(a)}_{j}-\partial_{j}e^{(a)}_{i})$ is the 
anholonomity object for the triad (\ref{ea}); for I-VII Bianchi types values
of its components numerically coincide with the corresponding structure 
constants $C^{a}{}_{bc}$. The list of explicit expressions for $\xi_{(a)},
e^{(a)}, C^{a}{}_{bc}, \hat{C}^{k}{}_{ij}$ for any Bianchi type is given in 
\cite{banach}.

We choose the constant matrix $\beta_{ab}$ in (\ref{ng}) to be positive
definite. This important condition generalises results of Maitra \cite{maitra},
and ensures the absence of closed time-like curves.

One can immediately see that space-times (\ref{met0}) admit, besides tree
Killing vector fields $\xi_{(a)}$, a nontrivial {\it conformal} Killing
vector
\begin{equation}
\xi_{\rm conf}=R\partial_t .\label{conf}
\end{equation}

All models in the class (\ref{met0}) have a number of common remarkable
properties:
\begin{itemize}
\item
Space-time manifolds are spatially homogeneous and completely causal.
\item
MBR is totally isotropic for any moment of $t$.
\item
Rotation (\ref{rot}) does not produce parallax effects.
\end{itemize}

These properties were proved in \cite{banach,jetp}, and here we only remark
that causality is provided by the positivity of $\beta_{ab}$ (one can
immediately write the original G\"odel metric (\ref{gd}) in the form 
(\ref{met0}) and check that its $\beta_{ab}$ matrix is not positive definite),
while isotropy of MBR and absence of parallax effects are related to the
existence of the conformal Killing vector (\ref{conf}). Hence, the most
strong limits on the cosmic rotation, obtained earlier from the study of
MBR \cite{haw,col,bjs} and of the parallaxes in rotating world \cite{tre,rub},
are not true for this class of cosmologies. 

Summarizing, cosmological models with rotation and expansion (\ref{met0})
solve the first three problems of cosmic rotation. This class of metrics
is rich enough, as it contains all kinds of worlds: open and closed with
different topologies.  

\section{G\"odel-type cosmological model}

To the end of this paper we will consider now the natural non-stationary
generalization of the original G\"odel metric (\ref{gd}) which has drawn
considerable attention in the literature. This generalized model is described
by the interval
\begin{equation}
ds^2 = dt^2 - 2\sqrt{\sigma}R(t)e^{mx}dtdy -
R^2 (t)(dx^2 + ke^{2mx}dy^2 + dz^2),\label{met1}
\end{equation}
where we denoted $x^1 =x, x^2 =y, x^3 = z$, and $m,\sigma,k > 0$ are
constant parameters. The condition $k > 0$ guarantees the absence of 
closed time-like curves. The metric (\ref{met1}) is usually called the 
{\it G\"odel--type model} with rotation and expansion. Coordinate $z$ gives
the direction of the global rotation, the magnitude of which
\begin{equation}
\omega = \sqrt{{1\over 2}\omega_{\mu\nu}\omega^{\mu\nu}}=
{m\over 2R}\sqrt{\sigma\over {k+\sigma}}\label{rotged}
\end{equation}
decreases in expanding world.

The three Killing vector fields are 
\begin{equation}
\xi_{(1)} ={1\over m}\partial_x -
y\partial_y ,\quad  \xi_{(2)} =\partial_y ,\quad  
\xi_{(3)} =\partial_z .\label{kil}
\end{equation}
These satisfy commutation relations
\begin{equation}
[\xi_{(1)},\xi_{(2)}]=\xi_{(2)},\quad
[\xi_{(1)},\xi_{(3)}]=[\xi_{(2)},\xi_{(3)}]=0,\label{comm}
\end{equation}
showing that the model (\ref{met1}) belongs to the Bianchi type III.

>From the point of view of the Petrov classification, one can verify that
the G\"odel--type model is of the type $D$. 

It is convenient to choose at any point of the space-time (\ref{met1}) a
local orthonormal (Lorentz) tetrad $h^{a}_{\mu}$ so that, as usual,
$g_{\mu\nu}=h^{a}_{\mu}h^{b}_{\nu}\eta_{ab}$ with $\eta_{ab}={\rm diag}(+1,
-1,-1,-1)$ the standard Minkowski metric. This choice is not unique, and
we will use the gauge in which
\begin{equation}
h^{\hat{0}}_{0}=1,\quad h^{\hat{0}}_{2}=-R\sqrt{\sigma}e^{mx},\quad
h^{\hat{1}}_{1}=h^{\hat{3}}_{3}=R,\quad h^{\hat{2}}_{2}=Re^{mx}\sqrt{k+\sigma}.
\label{tetrad}
\end{equation}
Hereafter a caret denotes tetrad indices; Latin alphabet is used for the 
local Lorentz frames, $a,b,...=0,1,2,3$.

\subsection{Dynamical realizations}

Several dynamical realizations (i.e. construction of exact cosmological
solutions for the gravitational field equations) of the G\"odel--type metric 
(\ref{met1}) are known. In the Einstein's general relativity theory such
models were described in \cite{krepan,kre,izv}, with different matter sources.
Rotation, spin and torsion are closely interrelated in the Poincar\'e gauge 
theory of gravity \cite{RMP,IS,PBO,IPS,MieB}, and hence it is quite natural 
to study cosmologies with rotation within the gauge gravity framework. 
General preliminary analysis of the separate stages of the universe's 
evolution was made in our works \cite{iko1,iko2,iko3}, while in 
\cite{ecsk,io} complete cosmological scenarios are considered. 

Here we shall mention a model \cite{io} in which the gravitational field
dynamics is determined by the minimal quadratic Poincar\'e gauge model. This
is perhaps the closest extension of the Einstein's general relativity theory. 
The matter source is represented by the Weyssenhoff spinning fluid and the 
magnetic field. In the analysis of the evolution of the scale factor $R(t)$ 
it is convenient to distinguish several qualitatively different stages in the 
history of the universe. First stage is the shortest one and it describes a 
bounce at $t=0$. There is no initial singularity due to the dominating spin 
contribution. Matter is characterized by the approximate equation of state 
$p\approx \varepsilon$ at this stage. Next comes the second stage when the 
scale factor increases like $\sqrt{t}$, while the equation of state is of 
the radiation type, $p\approx \varepsilon /3$. This expansion lasts until 
the size of the metagalaxy approaches $\approx 10^{27}$ cm. After this the 
``modern'' stage starts with the effectively dust equation of state 
$p_{0}\approx 0$. Scale factor still increases, but the deceleration of
expansion takes place. The final stage depends on the value of the 
cosmological term, and either the future evolution enters the eternal
de Sitter type expansion, or expansion ends and a contraction phase starts. 
The details of this complete scenario \cite{io} depend on the values of 
the coupling constants which determine the structure of the gravitational 
Lagrangian. The principal difficulty of this dynamical realization, in our 
opinion, is presented by the magnitude of magnetic field which at the 
``modern'' stage should be close to the upper limits established for the 
global magnetic field from astrophysical observations. 

\section{Null geodesics in the G\"odel-type model}

Practically all the information about the structure of the universe and about
the properties of astrophysical objects is obtained by an observer in the
form of different kinds of electromagnetic radiation. Thus, in order to be
able to make theoretical predictions and compare them with observations,
it is necessary to know the structure of null geodesics in the cosmological
model with rotation. All the models from the class (\ref{met0}) have
three Killing vectors and one conformal Killing vector field. Hence the
null geodesics equations 
\begin{equation}
k^{\mu}\nabla_{\mu}k^{\nu}=0,\quad\quad k^{\mu}k_{\mu}=0 \label{geo}
\end{equation}
(where $k^{\mu}={dx^{\mu}\over ds}$ is the tangent vector to a curve
$x^{\mu}(s)$ with an affine parameter $s$) have four first integrals,
\begin{equation}
q_0 =\xi^{\mu}_{\rm conf}k_{\mu}, \quad 
q_a = - \xi^{\mu}_{(a)}k_{\mu}, \quad a=1,2,3.\label{1int}
\end{equation}
Solving (\ref{1int}) with respect to $k^{\mu}$, one obtains a system of
ordinary first order nonlinear equations which can be straightforwardly 
integrated. Complete solution of the null geodesics equations in the 
G\"odel--type model is given in \cite{prep,banach}, and here we present 
only short description of null geodesics in (\ref{met1}). 

To begin with, let us define convenient parameterization of null geodesics.
Without loosing generality (using the spatial homogeneity) we assume that an
observer is located at the space-time point $P=(t=t_0 , x=0, y=0, z=0)$.
Now, arbitrary geodesics which passes through $P$ is naturally determined by
its initial direction in the local Lorentz frame of observer at this point.
In the tetrad (\ref{tetrad}) we may put
\begin{equation}
k^{a}_P =(h^{a}_{\mu}k^{\mu})_P = (1, \sin\theta\cos\phi, \sin\theta\sin\phi, 
\cos\theta ),\label{kini}
\end{equation}
where $\theta,\phi$ are standard spherical angles parameterizing the celestial 
sphere of an observer. Then from (\ref{conf}), (\ref{kil}), (\ref{kini}) and
(\ref{1int}) one finds the values of the integration constants
\begin{eqnarray}
q_0 &=& R_0,\nonumber\\
q_1 &=& {R_0\over m}\sin\theta\cos\phi,\nonumber\\
q_2 &=& R_0 (\sqrt{\sigma} + 
\sqrt{k+\sigma}\sin\theta\sin\phi),\label{const}\\
q_3 &=& R_0 \cos\theta.\nonumber
\end{eqnarray}
Generic null geodesics, initial directions of which satisfy 
$(\sin\theta\sin\phi +\sqrt{\sigma\over{k+\sigma}})\neq 0$, are then 
described by 
\begin{equation}
e^{-mx}={{\sqrt{\sigma} + \sqrt{k+\sigma}\sin\theta\sin\Phi}\over
{\sqrt{\sigma} + \sqrt{k+\sigma}\sin\theta\sin\phi}},\label{x}
\end{equation}
\begin{equation}
y={{\sin\theta (\cos\Phi - \cos\phi )}\over 
{m(\sqrt{\sigma} + \sqrt{k+\sigma}\sin\theta\sin\phi)}},\label{y}
\end{equation}
\begin{equation}
z=\left({{k+\sigma}\over k}\right)\cos\theta\left[\int^{t}_{t_0}{dt'\over
R(t')} + \sqrt{\sigma\over{k+\sigma}}\left({{\Phi - \phi}\over m}\right)
\right],\label{z}
\end{equation}
where the function $\Phi(t)$ satisfies the differential equation
\begin{equation}
{d\Phi\over dt}= - {m\over R}\left({
{\sqrt{\sigma\over {k+\sigma}} + \sin\theta\sin\Phi} \over
{1 + \sqrt{\sigma\over {k+\sigma}}\sin\theta\sin\Phi}}\right)\label{P}
\end{equation}
with initial condition $\Phi(t_0 )=\phi$. 

For a detailed discussion of rays which lie on the initial cone 
$(\sin\theta\sin\phi +\sqrt{\sigma\over{k+\sigma}})= 0$ and different 
subcases of (\ref{x})-(\ref{P}) see \cite{banach}. 

\section{Observations in rotating cosmologies}

The qualitative picture of specific rotational effects which could be 
observed in the G\"odel--type model (\ref{met1}) is in fact independent of
the dynamical behavior of the scale factor $R(t)$. To some extent the same
is true also for the quantitative estimates, especially if one uses the 
Kristian-Sachs formalism \cite{ks,mac} in which all the physical and 
geometrical observable quantities are expressed in terms of power series in 
the affine parameter $s$ or the red shift $Z$. In this case the description
of observable effects on not too large (although cosmological) scales 
involves only the modern values (i.e. calculated at the moment of observation
$t=t_0$) of the scale factor $R_{0}=R(t_{0})$, Hubble parameter $H_{0}=
(\dot{R}/R)_{P}$, rotation value $\omega_{0}=\omega(t_{0})$, deceleration 
parameter $q_{0}=-({\buildrel .. \over R}R^{2}/\dot{R}^{2})_P$, etc.  

In this section we will discuss possible observational manifestations of 
the cosmic rotation in the G\"odel--type universe. Estimates for the value
of vorticity and for the direction of rotation axis can be find from the
recent astrophysical data, see sects. 6.2 and 6.3. 

\subsection{Classical cosmological tests}

Classical cosmological tests, such as apparent magnitude -- red shift ($m-Z$),
number counts -- red shift ($N-Z$), angular size -- red shift relations, and
some other, reveal specific dependence of astrophysical observables on the 
angular coordinates ($\theta, \phi$) in a rotating world. Thus a careful 
analysis of the angular variations of empirical data over the whole celestial 
sphere is necessary. 

The knowledge of null geodesics enables one to obtain the explicit form of 
the area distance $r$ between an observer at a point $P$ and any source $S$,
which is a crucial step in deriving classical cosmological tests. The area
distance is defined \cite{ellis,ks,mac} by 
\begin{equation}
dA_S = r^2 d\Omega_P, \label{area}
\end{equation}
where $dA_S$ is the intrinsic area of the source which subtends the solid
angle $d\Omega_P$ at $P$ when observer looks at a source $S$. In general,
$r$ is thus a function of direction of observation, that is $r=r(\theta,\phi)$.
(Besides this, $r$ of course depends on the value of the affine parameter
$s$, or equivalently on the moment of $t$ at which source radiates a ray
detected by an observer at $t_0$).

Using (\ref{x})-(\ref{P}), one can find for the area distance along the axis
of rotation the following exact result
\begin{equation}
r^2(t; \theta=0) = {\sin^2\left(\int_{t_0}^{t}dt'\omega(t')\right)
\over \omega^{2}(t)}.\label{dist0}
\end{equation}
In general, rotational effects are always maximal in the directions close to
the $z$ axis, and quite remarkably observations in the direction of rotation
can be described by simple and clear formulas. As for an arbitrary direction,
exact formulas become very complicated and it is much more convenient to
replace them by the Kristian-Sachs expansions. Recall that the red shift $Z$,
which reflects the dependence of frequency on the motion of a source and
observer, is defined by
\begin{equation}
1+ Z = {(k^{\mu}u_{\mu})_S\over(k^{\mu}u_{\mu})_P},\label{red}
\end{equation}
where $u^{\mu}$ is the four--velocity of matter in the universe. In the
Kristian-Sachs approach this exact relation is replaced by the expansion 
\cite{ks,mac},
\begin{equation}
1 + Z = 1 + rK^{\mu}K^{\nu}(\nabla_{\mu}u_{\nu})_P +
{1\over 2}r^2 K^{\mu}K^{\nu}K^{\lambda}(\nabla_{\mu}\nabla_{\nu}
u_{\lambda})_P + ...,\label{zr}
\end{equation}
where
$$
K^{\mu}=\left({k^{\mu}\over k^{\nu}u_{\nu}}\right)_P .
$$
One can invert (\ref{zr}) and use the resulting expansions in the 
calculations of observable effects in rotating cosmologies. Now all the 
angular dependent rotational contributions are contained in the coefficients 
of these expansions. 

For the G\"odel--type cosmology (\ref{met1}) the classical ($m-Z$) and 
($N-Z$) relations read as follows.

{\sl Apparent magnitude $m$ vs. red shift $Z$}:
$$
m = M - 5\log_{10}H_{0} + 5\log_{10}Z + {5\over 2}(\log_{10}e)(1-q_{0})Z +
$$
$$
- 5\log_{10}\left(1 + \sqrt{\sigma\over{k+\sigma}}\sin\theta\sin\phi\right)+
$$
\begin{equation}
- {5\over 2}(\log_{10}e){\omega_0\over H_0}{\sin\theta\cos\phi
\left(\sqrt{\sigma\over{k+\sigma}} + \sin\theta\sin\phi\right)\over
\left(1 + \sqrt{\sigma\over{k+\sigma}}\sin\theta\sin\phi\right)^2} Z +
O(Z^2).\label{mz}
\end{equation}
{\sl Number of sources $N$ vs. red shift $S$}:
$$
{dN\over d\Omega} = {n_{0}Z^3 \over {3H_{0}^{3}\left(1 + 
\sqrt{\sigma\over{k+\sigma}}\sin\theta\sin\phi\right)^3}}\Bigg[ 1 -
{3\over 2}(1 + q_{0})Z -
$$
\begin{equation}
- 3{\omega_0 \over H_0}{\sin\theta\cos\phi
\left(\sqrt{\sigma\over{k+\sigma}} + \sin\theta\sin\phi\right)\over
\left(1 + \sqrt{\sigma\over{k+\sigma}}\sin\theta\sin\phi\right)^2} Z +
O(Z^2)\Bigg].\label{nz}
\end{equation}
In (\ref{mz})-(\ref{nz}) $M=-{5\over 2}\log_{10}L_{S}$ is the  absolute 
magnitude of a source with an intrinsic luminosity $L_S$ and $n_0$ is the
modern value of number density of sources $n=n(t)$ (as usual, (\ref{nz}) is
derived under the assumption of the absence of source evolution). 

The ($N-Z$) relation describes the number of sources observed in a solid
angle $d\Omega$ up to the value $Z$ of red shift. One can estimate the global 
difference of the number of sources visible in two hemispheres of the sky, 
$N^{+}, N^{-}$, by integrating (\ref{nz}). The result is
\begin{equation}
{{N^{+}- N^{-}}\over{N^{+}+ N^{-}}}= {1\over 2}\sqrt{\sigma\over{k+\sigma}}
\left(3 - {\sigma\over{k+\sigma}}\right) + O(Z^2 ).\label{NN}
\end{equation}
It seems worthwhile to draw attention to the absence of a correction 
proportional to $Z$ in (\ref{NN}). It is difficult to make a comparison of 
our results with \cite{wesson,mavr,fen} as they study stationary rotating 
models in which there is no red shift. 

For some time classical cosmological tests were carefully carried out for
standard models, but later it was recognized that evolution of physical
properties of sources often dominates over geometrical effects. However,
specific angular irregularities predicted in rotating cosmologies, (\ref{mz}),
(\ref{nz})-(\ref{NN}), may revive the importance of the classical tests. 
     
\subsection{Periodic structure of the universe}

Recent analysis of the large--scale distribution of galaxies \cite{bro}
has revealed an apparently periodic structure of the number of sources as
a function of the red shift. Cosmic rotation may give a natural explanation
of this fact \cite{per}. The crucial point is the structure of null 
geodesics in the G\"odel--type model: explicit solutions (\ref{x})-(\ref{z}) 
demonstrate a helicoidal behavior of rays in directions close to the 
rotation axis. This yields a periodicity of the area distance as a function 
of red shift, and hence the visible distribution of sources turn out to be 
also approximately periodical in $Z$. 

This effect is most transparent for the direction of rays straight along
the axis of rotation. The area distance is then given by (\ref{dist0}). In
order to be able to make some quantitative estimates, let us assume the
polynomial law for the scale factor,
\begin{equation}
R(t)=R_{0}\left({t-t_\infty\over{t_0 - t_\infty}}\right)^b ,\label{scale}
\end{equation}
which is naturally arising in a number of cosmological scenarios ($0<b<1$). 
Then it is straightforward to derive (analogously to (\ref{nz})) the 
distribution of number of sources per red shift per solid angle,
\begin{equation}
{dN\over{d\Omega dZ}}={n_{0}\over{\omega_{0}^{2}H_{0}(1+Z)^{1/b}}}
\sin^2 \left({b\omega_{0}\over{(1-b)H_{0}}}\left[(1+Z)^{(b-1)/b} - 
1\right]\right).\label{pereff}
\end{equation}
This shows that the apparent distribution of visible sources is an 
oscillating function of red shift, with slowly decreasing amplitude. Similar 
generalized formula can be obtained for arbitrary directions, so that 
(\ref{pereff}) is modified by additional angular dependence of the magnitude 
of successive extrema of distribution function. 

Observational data \cite{bro} give for the distance between maxima the value
$128 h^{-1}$ Mps (where $H_{0}=100 h {\rm km\ sec}^{-1}{\rm Mps}{-1}$). From
this one can estimate the rotation velocity which is necessary to produce
such a periodicity effect,
\begin{equation}
\omega_{0}\approx 74 H_{0}.\label{omper}
\end{equation}
This result does not depend on $b$.

\subsection{Polarization effect}

Cosmic rotation affects polarization of radiation which propagates in 
(\ref{met1}), and this produces a new observable effect which has been 
already reported in the literature by Birch \cite{birch1,birch2}. In the 
geometrical optics approximation, polarization is described by a space-like
vector $f^{\mu}$ which is orthogonal to the wave vector, $f_{\mu}k^{\mu}=0$,
and is parallelly transported along the light ray, 
\begin{equation}
k^{\mu}\nabla_{\mu}f^{\nu}=0.\label{pol}
\end{equation}
Study of (\ref{pol}) reveals that the cosmic rotation forces a polarization 
vector to change its orientation during propagation along the null geodesics.
It is clear that this conclusion has physical meaning only when one defines
a frame at any point of the ray with respect to which polarization rotates. 
Let us describe how this can be achieved.

As it is well known, gravitational field affects the properties of an image 
of a source, such as shape, size and orientation \cite{sachs,nf}. Like the 
rotation of polarization vector, deformation and rotation of image depend on 
local coordinates and on the choice of an observer's frame of reference. 
However, one can consider the combination of two problems, and this gives 
rise to a truly observable effect which is coordinate and frame independent. 
Putting it in another way, one should calculate the influence of the cosmic 
rotation on {\it the relative angle} $\eta$ between the polarization vector 
and the direction of a major axis of an image. [This problem is discussed in
the recent paper \cite{panov}, but it is incorrect, in our opinion]. 

Most conveniently this can be done within the framework of the Newman-Penrose
spin coefficient formalism. Namely, it is enough to construct a null frame 
$\{l,n,m,\overline{m}\}$ is such a way that $l$ coincides with the wave 
vector $k$, and the rest of the vectors are covariantly constant along $l$.
Then we can consider $m$ as a polarization vector, and thus deformation of 
an image of a source, calculated with respect to this frame $\{l,n,m,
\overline{m}\}$, gives at the end the observable relative angle $\eta$.
Let us describe explicitly the null frame:
$$
l = {R_0\over R}\Bigg[\left\{1 + 
\sqrt{\sigma\over {k+\sigma}}\sin\theta\sin\Phi\right\}\partial_t +
$$
\begin{equation}
+{1\over R}\sin\theta\cos\Phi\partial_x +
{e^{-mx}\over R\sqrt{k+\sigma}}\sin\theta\sin\Phi\partial_y +
{1\over R}\cos\theta\partial_z\Bigg],\label{ll}
\end{equation}
\begin{equation}
n = {R\over R_0}\left({1\over{1+\cos\theta}}\right)
\left[\partial_t - {1\over R}\partial_z\right],\label{nn}
\end{equation}
$$
m ={e^{i\Psi}\over \sqrt{2}}\Bigg[ \left\{\sqrt{\sigma\over {k+\sigma}} +
i\left({\sin\theta\over{1+\cos\theta}}\right)e^{-i\Phi}\right\}\partial_t +
$$
\begin{equation}
+{i\over R}\partial_x + {e^{-mx}\over R\sqrt{k+\sigma}}\partial_y -
{i\over R}\left({\sin\theta\over{1+\cos\theta}}\right)
e^{-i\Phi}\partial_z\Bigg],\label{m1}
\end{equation}
$$
\overline{m}={e^{-i\Psi}\over \sqrt{2}}\Bigg[\left\{
\sqrt{\sigma\over {k+\sigma}} -
i\left({\sin\theta\over{1+\cos\theta}}\right)e^{i\Phi}\right\}\partial_t -
$$
\begin{equation}
- {i\over R}\partial_x + {e^{-mx}\over R\sqrt{k+\sigma}}\partial_y +
{i\over R}\left({\sin\theta\over{1+\cos\theta}}\right)
e^{i\Phi}\partial_z\Bigg].\label{m2}
\end{equation}
Here 
\begin{equation}
\Psi(t,z)=z{m\over 2}\sqrt{\sigma\over {k+\sigma}} + \Phi(t).\label{Psi}
\end{equation}
One should note that (\ref{ll})-(\ref{m2}) is a smooth field of frames which 
cover all the space-time manifold. Direct computation proves that (\ref{ll})
is the null geodesics congruence with an affine parameterization. We say
that this congruence is oriented along the direction given by the spherical
angles $(\theta,\phi)$ in the local Lorentz frame of an observer at $P$, 
because a null geodesics (\ref{x})-(\ref{P}) belongs to this congruence.

Direct computation of the spin coefficients (we are using definitions of 
\cite{chandra}, and denote spin coefficients by tildes in order to 
distinguish them from other quantities in this paper) gives
\begin{equation}
\widetilde{\varepsilon} = 0, \quad\quad \widetilde{\kappa} = 0,\label{ep-kap}
\end{equation}
\begin{equation}
\widetilde{\lambda} = 0, \quad\quad \widetilde{\nu} = 0,\label{la-nu}
\end{equation}
$$
\widetilde{\rho} = - {R_0\over R^2}\Bigg[\dot{R}\left\{1 + 
\sqrt{\sigma\over {k+\sigma}}\sin\theta\sin\Phi \right\} +
$$
$$
+{m\over 2}\left({k\over{k+\sigma}}\right) {\sin\theta\cos\Phi\over
(1 + \sqrt{\sigma\over {k+\sigma}}\sin\theta\sin\Phi )}\Bigg] +
$$
\begin{equation}
+ i{R_0\over R^2}{m\over 2}\cos\theta\left(
{{\sqrt{\sigma\over {k+\sigma}} + \sin\theta\sin\Phi} \over
{1 + \sqrt{\sigma\over {k+\sigma}}\sin\theta\sin\Phi}}\right),\label{rho}
\end{equation}
$$
\widetilde{\sigma} ={R_0\over R^2}{m\over 2}\left({k\over{k+\sigma}}\right)
{e^{2i\Psi}\sin\theta \over {(1 + 
\sqrt{\sigma\over {k+\sigma}}\sin\theta\sin\Phi )}}
\Big(-\cos\Phi\Big[2\cos\theta - 1 - 
$$
\begin{equation}
- 2\cos^{2}\Phi (\cos\theta -1)\Big] +
i\sin\Phi\left[\cos\theta - 2\cos^{2}\Phi (\cos\theta -1)\right]\Big),
\label{sigma}
\end{equation}
and we do not write other spin coefficients, because their values are
irrelevant. Only one important step is to be done: spin coefficient 
$\widetilde{\pi}\neq 0$ in the frame (\ref{ll})-(\ref{m2}), and we need to
make an additional Lorentz transformation
\begin{equation}
l\longrightarrow l,\quad n\longrightarrow n + a^{*}m + a\overline{m} +
a^{*}a\ l, \quad m\longrightarrow m + al, \quad \overline{m}\longrightarrow
\overline{m} + a^{*}l,\label{add}
\end{equation}
where the function $a$ satisfies equation $l^{\mu}\partial_{\mu}a +
\widetilde{\pi}^{*}=0$. This ensures that $\widetilde{\pi}=0$ in a new 
frame, while remarkably the transformation (\ref{add}) does not change any
of the spin coefficients (\ref{ep-kap})-(\ref{sigma}). Thus one obtains
finally the field of null frames $\{ l,n,m,\overline{m}\}$ with the required
properties: $l$ is the null geodesics congruence with affine parameterization,
while $n,m,\overline{m}$ are covariantly constant along $l$. The latter is
equivalent to $\widetilde{\kappa}=\widetilde{\varepsilon}=\widetilde{\pi}=0$.

As it is well known, deformation and rotation of an image along a null 
geodesics are described by the optical scalars
\begin{equation}
\widetilde{\vartheta}=-{\rm Re}\widetilde{\rho},\quad
\widetilde{\omega}={\rm Im}\widetilde{\rho},\quad \widetilde{\sigma}.
\end{equation}
We now assume, for definiteness, that the polarization vector $f^\mu$ 
coincides with the vector $m^\mu$ of the above null frame. Let the image of
a source, as seen at the point corresponding to the value $s=s_1$ of the
affine parameter, be an ellipse with the major axes $a$ and minor axis $b$. 
Then one can straightforwardly obtain for the angle of rotation of the major
axis of the image at $s_2 =s_1 + \delta s$,
\begin{equation}
\delta\eta = -\widetilde{\omega}\delta s - 
{{a^2 + b^2}\over{a^2 - b^2}}{\rm Im}\widetilde{\sigma}\delta s.\label{eta}
\end{equation}
Integration along a ray gives finite angle of rotation.  

It is worthwhile to notice that along the cosmic rotation axis the observer
at $P$ finds for the optical scalars
\begin{equation}
\widetilde{\vartheta}_{P}=H_{0},\quad \widetilde{\omega}_{P}=\omega_{0},\quad
\widetilde{\sigma}=0,
\end{equation}
thus the effect of rotation of the polarization vector in this direction is
most explicit.

As for an arbitrary direction of observation, with the help of the 
Kristian--Sachs approach we finally find from (\ref{eta})
\begin{equation}
\eta = \omega_0 \ r \ \cos\theta + O(Z^{2}).\label{final}
\end{equation}
This result is in good agreement with the observational data reported 
\cite{birch1} on the dipole anisotropy of distribution of the difference 
between the position angles of elongation (the major axis) and polarization 
in a sample of 3CR radiosources. The estimate for the direction and the
magnitude of cosmic rotation, obtained from the Birch's data \cite{jetp,grg},
read
\begin{equation}
l^{\circ}=295^{\circ}\pm 25^{\circ},\quad\quad 
b^{\circ}=24^{\circ}\pm 20^{\circ},\label{dir}
\end{equation}
\begin{equation}
\omega_{0}=(1.8\pm 0.8)H_{0}.\label{om2}
\end{equation}

\section{Conclusions}

In our discussion of the properties of rotating cosmologies we have paid 
special attention to the G\"odel--type model (\ref{met1}). However the main
conclusions are true also for the whole class of metrics (\ref{met0}). A
series of papers is now under preparation in which we make exact estimates
for the cosmic rotation effects, and present also their dynamical 
realizations, for all nine Bianchi type rotating cosmologies. As one can 
notice, in some cases the above numerical estimates for the value of 
vorticity do not agree e.g., (\ref{omper}) and (\ref{om2}). One can only 
remark in this relation that too few empirical data were analysed until now, 
and further detailed discussion is required in order to make final estimates. 
It may also turn out that some of the above mentioned effects are explained 
after all by different physical (and geometrical) reasons, not related to 
the cosmic rotation. In the light of the modern COBE results 
\cite{co1,co2,co3} the purely rotating models (\ref{met0}) should be 
replaced by cosmologies with nontrivial shear. Preliminary analysis of such 
generalizations shows that rotating models can be made compatible with the 
COBE data without destroying the rest of the rotational effects (in 
particular, without essential modification of the polarization rotation 
formulas).

We believe that the cosmic rotation is an important physical effect which
should find its final place in cosmology. In this paper we outlined one
of the possible theoretical frameworks which can underlie our understanding
of this phenomenon.

\bigskip
{\bf Acknowledgments}. YNO is grateful to Friedrich W. Hehl for useful
discussions. The work of YNO was partly supported by the Alexander von
Humboldt Foundation and Deutsche Forschungsgemeinschaft (Bonn) grant 
He 528/17-1.

\end{document}